\documentclass[11pt,a4paper]{article} 
\usepackage{jheppub}

\newcommand{\be}{\begin{equation}}
\newcommand{\ee}{\end{equation}}
\newcommand{\beqa}{\begin{eqnarray}}
\newcommand{\eeqa}{\end{eqnarray}}

\newcommand{\tr}{^{3\!}R}

\usepackage{graphicx}
\usepackage{amsmath}
\usepackage{amssymb}

\title{Inflation with stable anisotropic hair: \\ is it cosmologically viable?}
\author[a]{Sigbj{\o}rn Hervik,} \author[b]{David F.~Mota} \author[b]{and Mikjel Thorsrud}
\affiliation[a]{Faculty of Science and Technology, University of Stavanger, N-4036 Stavanger, Norway} 
\affiliation[b]{Institute of Theoretical Astrophysics, University of Oslo, N-0315 Oslo, Norway}
\emailAdd{sigbjorn.hervik@uis.no} 
\emailAdd{d.f.mota@astro.uio.no} 
\emailAdd{mikjel.thorsrud@astro.uio.no}
\abstract{Recently an inflationary model with a vector field coupled to
  the inflaton was proposed and the phenomenology studied for the
  Bianchi type I spacetime. It was found that the model demonstrates a
  counter-example to the cosmic no-hair theorem since there exists a
  stable anisotropically inflationary fix-point. One of the
  great triumphs of inflation, however, is that it explains the
  observed flatness and isotropy of the universe today without
  requiring special initial conditions.  Any acceptable model for
  inflation should thus explain these observations in a satisfactory
  way. To check whether the model meets this
  requirement, we  introduce curvature to the background geometry and consider
  axisymmetric spacetimes of Bianchi type II,III and the
  Kantowski-Sachs metric.  We show that the anisotropic Bianchi type I
  fix-point is an attractor for the entire family of such
  spacetimes.  The model is
  predictive in the sense that the universe gets close to this
  fix-point after a few e-folds for a wide range of initial
  conditions. If inflation lasts for $N$ e-folds, the
  curvature at the end of inflation is typically of order $\sim
  e^{-2N}$. The anisotropy in the expansion rate at the end of inflation, on the other
  hand, while being small on the one-percent level, is highly significant.  We show that after
  the end of inflation there will be a period of isotropization
  lasting for $\sim\frac{2}{3}N$ e-folds. After that the shear scales as the curvature and
  becomes dominant around $N$ e-folds after the end of inflation. For
  plausible bounds on the reheat temperature the minimum
  number of e-folds during inflation, required for consistency with
  the isotropy of the supernova Ia data, lays in the interval ($21,48$). Thus the results obtained for our restricted class of spacetimes
  indicates that inflation with anisotropic hair is
  cosmologically viable.}
\arxivnumber{1109.3456}
\begin{document} 
\maketitle
\flushbottom

\section{Introduction}
It is plausible that inflation occurred around the energy scale of the grand unification. Since we do not have direct experimental access to such energy scales we need to be open-minded regarding the physics possibly occurring there. Indeed, when theoretical models of inflation are compared to observations they might provide the clue to understand fundamental physics at non-accessible energies.  In this paper we are concerned with a specific model that violates isotropy; i.e. three-dimensional rotational invariance. Although rotational invariance is a well established feature of low-energy physics, several puzzling features of the large scale CMB anisotropies hints that this symmetry might have been broken during inflation \cite{Efstathiou04,oliveira04,copi04,eriksen04,hansen04,nico1}. At the same time recent progress in theoretical models has clarified that inflation with anisotropic hair is a theoretical possibility \cite{dav4,hervik06, hervik06nr2, carroll07, golovnev08, kanno08, dav1,dav3,hervik10,nico2}.  It has become clear, however, that specific realizations are often plagued by instabilities, either ghosts or unstable growth of the linearized perturbations \cite{dav2,himmetoglu09,himmetoglu08,carroll09}.

In a series of recent papers M. Watanabe, S. Kanno and J. Soda studied the cosmology of a model with anisotropic hair which, apparently, is free of instabilities \cite{soda09,soda09mag,soda10,soda10pert,soda10cmb} (for related works by other authors also see \cite{emami,emami2,dimo,dimo2,do}). Their model is inspired from supergravity which includes a massless vector field coupled to the scalar field(s). The vector part of the supergravity action has so far been neglected in cosmology, but for an approprately chosen coupling, the authors demonstrated interesting cosmological implications when the backreaction to geometry is properly accounted for.  In particular, it is shown that the anisotropy in the expansion rate is stable and proportional to the slow roll parameter \cite{soda09}. Even in the case when the anisotropy is very small, say on the micro level, the statistical anisotropy imprinted in primordial fluctuations can be significant \cite{soda10pert}. Thus, it has become clear that, in high precision cosmology, one cannot always neglect the back-reaction of vector-fields to geometry.

One of the essential features of the conventional inflationary scenarios, however, is that they explain the observed flatness and isotropy of the universe today without requiring special initial conditions. For spatially homogeneous models of non-positive curvature containing comoving fluids obeying the strong energy condition, this is a consequence of the \emph{cosmic no-hair theorem} \cite{wald, moss, nohairpowerlaw} which guarantees that the curvature and shear of the spatial geometry decay rapidly during inflation. Given that inflation is stable for a sufficiently large time, the universe will, in agreement with observations, remain almost flat and isotropic until today.\footnote{As a rule of thumb, 60 e-folds is sufficient.}  The above mentioned papers, however, clearly demonstrate that the theorem may be violated if a massless vector field is appropriately coupled to the scalar field (and necessarily violating the assumptions given in the theorem). The phenomenology of the model has so far only been studied in the context of the spatially flat Bianchi type I model. The implications of curvature are therefore, until now, unexplored.  A crucial question is whether the model, while violating the cosmic no-hair theorem, is still consistent with the observed isotropy and flatness of the universe today. It is therefore important to understand the behavior of the model with more general initial conditions.  In this paper we shall introduce curvature to the model by considering axisymmetric spacetimes of Bianchi type II,III and the Kantowski-Sachs metric. To be specific we shall consider the well motivated case where both the potential of the scalar field and the coupling function between the vector and scalar fields are exponentials (this specific model is analyzed without spatial curvature in \cite{soda10}).

The organization of the paper is as follows.  In section \ref{chmodel} we give a brief introduction to the model, while in section \ref{chgeometry} we motivate our class of considered spacetimes. In section \ref{cheq} we derive the field equations.  Section \ref{chdsa1} is devoted to dynamical system analyses.  First, in \ref{chdsa2}, we characterize the phase-space by identifying and classifying all fix-points.  Then, in \ref{chdsa3}, we study the phase flow with arbitrary initial conditions.  As we shall see the model is predictive in the sense that it provides unambiguous initial conditions for the post-inflationary era. Finally, in section \ref{chlate}, we show that these initial conditions provide a viable cosmological scenario.

\section{The model \label{chmodel}}
The  starting point of the model is the following action for the metric $g_{\mu\nu}$, scalar field $\phi$ and vector field $A_\mu$:
\begin{equation}
	S = \int d^4x \sqrt{-g} \left[\frac{M_p^2}{2}R - \frac{1}{2}(\nabla \phi)^2 - V(\phi) - \frac{1}{4} f^2(\phi)F_{\mu\nu}F^{\mu\nu} \right],
\label{action}
\end{equation}
where $(\nabla \phi)^2=g^{\mu\nu}\nabla_\mu \phi \nabla_\nu \phi$ is the kinetic term of the scalar field, and $F_{\mu\nu}=\nabla_\mu A_\nu - \nabla_\nu A_\mu$ is the field strength of the vector field.  In agreement with conventional notation $g$ is the determinant of the metric, $R$ is the Ricci scalar and $M_p$ is the reduced Planck mass. We shall use metric signature $(-1,1,1,1)$. Motivated by dimensional reduction of higher dimensional theories we take the potential of the scalar field and the coupling functions to be exponentials:
\begin{align}
V(\phi)&=V_0 e^{\lambda\frac{\phi}{M_p}},\\
f(\phi)&=f_0 e^{Q\frac{\phi}{M_p}},
\end{align}
where $\lambda$ and $Q$ are constant parameters that characterize the model. We shall assume slow-roll inflation which implies $\lambda\ll 1$.  We shall treat the coupling constant $Q$ as a free parameter, and study the implications on inflation for different values.  We consider a positive potential which implies $V_0>0$. There are no restrictions on the constant $f_0$. 
Variation with respect to $g^{\mu\nu}$, $\phi$ and $A_\mu$, respectively, gives the equations of motion:
\begin{align}
M_p^2 E^\mu_{\;\nu} &= {T}^{\mu(\phi)}_{\;\nu} + {T}^{\mu(A)}_{\;\nu},  \label{eqm1} \\
\nabla_\mu T^{\mu(\phi)}_{\;\nu} &=  -Q \frac{2\mathcal L_A}{M_p} \nabla_\nu \phi  ,  \label{eqm2} \\
\nabla_\mu T^{\mu(A)}_{\;\nu} &= + Q \frac{2\mathcal L_A}{M_p} \nabla_\nu \phi,  \label{eqm3} 
\end{align}
where $E^\mu_{\;\nu}$ is the Einstein tensor, and $\mathcal{L}_A=- \frac{1}{4} f^2(\phi)F_{\mu\nu}F^{\mu\nu}$ is the Lagrangian for the vector field.  Notice that the total energy-momentum tensor is conserved $\nabla_\mu(T^{\mu(\phi)}_{\;\nu}+T^{\mu(A)}_{\;\nu})=0$.  It is manifest from the equations of motions that the coupling leads to exchange of energy and momentum between the scalar field and the vector field. The rate is determined by the coupling constant $Q$.  It is the vector field that sources the shear degree of freedom of the expansion rate in this model.\footnote{Since we do not consider any shear  in the matter source, the term \emph{shear} will unambiguously refer to the geometrical freedom in the expansion rate.} Without the coupling to the scalar field, however, the vector field would decay rapidly leading to isotropization.\footnote{As an interesting digression we mention that in a model where inflation is driven by non-Abelian gauge vector fields (without any scalar field), the anisotropic hair vanish exponentially fast although the field is sufficiently stable to drive inflation, see \cite{azadeh} and references therein.}  

Finally we write down the components of the energy-momentum tensors for the scalar field and the vector field:
\begin{align}
{T}^{\mu(\phi)}_{\;\nu} &= - \delta^\mu_\nu\left( \frac{1}{2} (\nabla \phi)^2 +V(\phi)\right) + \nabla^\mu \phi \nabla_\nu \phi     \label{emt1} \\ 
{T}^{\mu(A)}_{\;\nu} &= - f^2(\phi) F^{\mu}_{\;\alpha}F^{\alpha}_{\; \nu} - \frac{1}{4}f^2(\phi) \delta^\mu_\nu F_{\alpha\beta}F^{\alpha\beta} \label{emt2}, 
\end{align}
where $\delta^\mu_\nu\!=\!g^{\mu\alpha}g_{\alpha\nu}$.

\section{On axisymmetric geometries \label{chgeometry}}
The usual approach for anisotropic models where the shear is sourced by a homogeneous vector field, is to assume axisymmetric geometries, often called local rotational symmetric (LRS) spacetimes.  The vector field is assumed to be aligned parallel with the axis of symmetry, ie. orthogonal to the plane of rotational symmetry. Although this is a widely used assumption in the literature \cite{carroll07,soda09,soda09mag,soda10,soda10pert,boehmer}, it seems like a proper discussion is lacking.\footnote{Ref. \cite{soda10pert} comments that the shear variable (in the considered two dimensional hypersurface) is exponentially decaying in an expanding universe. But the physical interesting quantity is the \emph{dimensionless} shear degree of freedom which, roughly speaking, measures the fraction between the anisotropic and isotropic parts of the expansion rate. As we shall see, this one is not necessarily exponentially decaying in an expanding universe.} Beside being a simplifying assumption we think that the approach is appealing since, on the background level, one has the same symmetry in the spacetime geometry as in the (total) matter field.  One can also loosely argue that if the symmetry was broken initially, spacetime would quickly become axisymmetric again, since there is no way to source a preferred direction in the plane orthogonal to the vector field.\footnote{Note that in quadratic theories the curvature, in some sense, may source itself creating the possibility for non-axisymmetric Bianchi type I solutions \cite{hervik06nr2}.} The purpose of this section is to show rigorously that this is actually what happens in the Bianchi type I spacetime.  We find that the universe isotropizes in the plane orthogonal to the vector field and that in the case of inflation this is a very rapid process.

We consider the most general Bianchi type I geometry:
\be
ds^2 = -dt^2 + a_x^2 dx^2 + a_y^2 dy^2 + a_z^2 dz^2,
\label{b1general}
\ee
where $a_x(t)$, $a_y(t)$ and $a_z(t)$ are three independent scale factors.  It is convenient to introduce three functions $\alpha(t)$, $\sigma_1(t)$ and $\sigma_2(t)$ defined by  
\be
a_x = e^{\alpha}e^{-\sigma_1-\sigma_2},\quad a_y = e^{\alpha}e^{\sigma_1}, \quad a_z = e^{\alpha}e^{\sigma_2}.
\ee
In the special case $\sigma_1(t)=\sigma_2(t)$ we recover the axisymmetric geometry usually assumed. The three Hubble factors $H_i\equiv \frac{\dot a_i}{a_i}$ can then be written:
\be
H_x = \dot\alpha-\dot\sigma_1-\dot\sigma_2, \quad H_y = \dot\alpha+\dot\sigma_1, \quad H_z = \dot\alpha+\dot\sigma_2,
\ee
where a dot denotes differentiation with respect to the cosmic time $t$. It is also useful to introduce a mean Hubble rate
\be
H\equiv \frac{1}{3} (H_x+H_y+H_z) = \dot\alpha.
\ee
As in the Friedman-Lema\^{i}tre-Robertson-Walker  (FLRW) metric, acceleration can be quantified in terms of the deceleration parameter $q$ defined by:  
\be
q=-1-\frac{\dot H}{H^2}.
\label{q}
\ee
Acceleration is then defined as $q<0$.  Note that $q<0$ is equivalent to $\frac{d^2}{dt^2} e^{\alpha}>0$. The function $e^{\alpha}$ can be interpreted as an isotropic scale factor. More precisely it is the geometric mean of the scale factors.  Thus our definition of acceleration is equivalent to acceleration of the geometric mean of the scale factors. These definitions are standard in homogenous cosmologies, and generalize the definitions of the FLRW model in a natural way.   

To proceed we impose the metric (\ref{b1general}) on the gravitational equation (\ref{eqm1}). We assume a single electric-type field. Since there are no spatial symmetries, apart from homogeneity of course, in the considered spacetime, we can without loss of generality align the field in the $x$-direction: $\mathbf F= \frac{1}{2}F_{\mu\nu}(t) dx^{\mu}\!\!\wedge\! dx^{\nu}=F_{10}(t)dx\!\wedge\! dt$. In the gauge $A_0=0$ this field configuration corresponds to the vector potential $\mathbf A = A_\mu dx^\mu= A_x(t) dx$ (where $A_x$ are related to $F_{10}$ by $F_{10}=-\dot A_x$). By taking linear combinations of the components of (\ref{eqm1}) one can eliminate the matter sources leading to the pure geometrical equation:
\be
(\ddot\sigma_2 - \ddot\sigma_1) = -3\dot\alpha (\dot\sigma_2 - \dot\sigma_1).
\label{physics}
\ee
We now introduce a shear degree of freedom:
\be
X_\perp \equiv \frac{H_z-H_y}{H} = \frac{\dot\sigma_2 - \dot\sigma_1}{\dot \alpha}.
\ee
The quantity $|X_\perp|$ is a measure of the anisotropy in the plane orthogonal to the vector field (as indicated by the subscript). For $|X_\perp| = 0$ the hypersurface is isotropic and we have the axisymmetric metric usually assumed.  Equation (\ref{physics}) then implies:
\be
\frac{d|X_\perp|}{d\alpha} = -|X_\perp|(2-q).
\label{isotropization}
\ee
In this equation we have changed to the number of e-folds, $\alpha$, as time parameter by the identity $\frac{d\alpha}{dt}=H$. From the gravitational equations it is easy to verify that $q< 2$ since we consider a positive potential $V(\phi)>0$. This implies $\frac{d|X_\perp|}{d\alpha}< 0$.  Thus we have shown that $|X_\perp|$ decays monotonically, although, in general, it will not necessarily decay rapidly.  In the case of inflation, however, we have by definition $q<0$. From (\ref{isotropization}) it then follows that $\frac{d|X_\perp|}{d\alpha}<-2|X_\perp|$.  Thus inflation guarantees that $|X_\perp|(\alpha)$ decays faster than the function $e^{-2\alpha}$.  To summarize we have showed that the Bianchi type I spacetime isotropizes in the plane orthogonal to the vector field, and that in the case of inflation this is a very rapid (exponential) process.       

Let us finally use this result to prove that there will be no inflationary fix points with anisotropic expansion in the considered hypersurface. A dynamical system analysis for the spacetime (\ref{b1general}) would require two shear variables, say $X_1= \frac{\dot\sigma_1}{\dot\alpha}$ and $X_2= \frac{\dot\sigma_2}{\dot\alpha}$.  Note that from the definition of $X_\perp$ we have $X_\perp=X_2 - X_1$. It follows that for a fix-point ($\frac{dX_1}{d\alpha}=\frac{dX_2}{d\alpha}=0$), we must have $\frac{dX_\perp}{d\alpha}=0$.  But, as shown above, this is not possible in our model. Thus there will be no additional fix-points if introducing a new shear degree of freedom. 

Although this analysis has been restricted to the Bianchi type I metric, we believe it sheds some light on the more general class of homogenous geometries, and we will use this as a motivation for the class of spacetimes to be introduced in the next section.

\section{Field equation for a class of homogenous and axisymmetric spacetimes \label{cheq}}

We shall now impose a class of homogenous spacetimes to the field equations. Motivated by the discussion above we consider a class of axisymmetric versions of Bianchi types I, II, III and the Kantowski-Sachs metric. We shall refer to these as BI, BII, BIII and KS, respectively. Type BI is spatially flat while BII, BIII and KS have anisotropic curvature in addition to the shear. For these spacetimes there is an intrinsic rotational symmetry in the spatial curvature which is aligned with the rotational symmetry of the expansion rate. Our considered class of spacetimes can be written on the form:
\begin{equation}
ds^2 = -dt^2 + e^{2\alpha(t)}\left( e^{-4\sigma(t)} w^1\!\!\otimes\! w^1 + e^{2\sigma(t)} w^2\!\!\otimes\! w^2 + e^{2\sigma(t)} w^3\!\!\otimes\! w^3 \right),   
\label{lineelement}
\end{equation}
where $w^i$ are three time-independent and mutually orthogonal one-forms which are related to the coordinate basis in table \ref{taboneforms}. These spacetimes have rotational symmetry in the plane spanned by $w^2$ and $w^3$. The rotational symmetry in the expansion rate is manifest from the line-element on this form. The rotational symmetry of the anisotropic curvature, on the other hand, is manifest first after an appropriate coordinate transformation.\footnote{As an example, in the Bianchi type II one can use $y=r\cos (\phi)$ and $z=r\sin(\phi)$.} To obtain the same symmetry in the (total) matter as in the spacetime geometry, we shall align the vector field parallel to $w^1$.

\begin{table}[h]
\newcommand\T{\rule{0pt}{2.6ex}}
\newcommand\B{\rule[-1.2ex]{0pt}{0pt}}
\begin{center}
\begin{tabular}{ l l l l l }     
\hline\hline                    
Spacetime \T\B & $\tr$ &  $w^1$ & $w^2$ & $w^3$  \\
  \hline
 BI \T & 0 & $dx$ & $dy$ & $dz$    \\
 BII & $-2k^2e^{-2\alpha-8\sigma}$ & $dx\!+\!k(y dz\!-\!zdy)$ & $dy$ & $dz$  \\
 BIII \B $\&$ KS & $2ke^{-2\alpha-2\sigma}$ & $dx$ & $(1\!-\!ky^2)^{-1/2}dy$ & $ydz$  \\
 \hline\hline   
\end{tabular}
\caption{Geometric variables in the various spacetimes. $w^i$ are the one-forms in the line element (\ref{lineelement}). $\tr$ is the spatial Ricci scalar. $k$ is a constant. For BIII $k<0$, while $k>0$  for KS.}
\label{taboneforms}
\end{center}
\end{table}

Since the one-forms are time-independent one can read the scale factors directly from the line-element (\ref{lineelement}):
\be
a_\parallel = e^{\alpha-2\sigma}, \quad a_\perp = e^{\alpha+\sigma},
\ee
where $a_\parallel$ and $a_\perp$ are the scale factors in the direction parallel and perpendicular to the vector field, respectively. The corresponding Hubble factors are:
\be
H_\parallel = \dot\alpha - 2\dot\sigma, \quad H_\perp = \dot\alpha + \dot\sigma.
\ee
The mean Hubble rate becomes similar as in the previous section, $H=\dot\alpha$.   

We shall now introduce two spacetime dependent coefficients $s_1$ and
$s_2$ with values specified in table \ref{tabsi}. The former is
defined by $\dot\tr = -2(\dot\alpha+s_1\dot\sigma)\tr$, where $\tr$ is the
three-dimensional Ricci scalar of constant time hypersurfaces. The
functions $\tr$ for the various spacetimes are specified in table
\ref{taboneforms}. The latter ($s_2$) determines the strength of a
coupling between the curvature and the energy density of the vector
field.\footnote{As we shall see, it turns out that such a coupling
  exists in BII, but not in BIII or KS.} We can then treat the entire
class of spacetimes in a unified way by expressing the equations in
terms of $s_1$,$s_2$ and $\tr$. 

Next, let us consider the matter sources. The most general field strength compatible with the class of spacetimes is on the form $\mathbf F =  \dot v \; w^0\!\!\wedge\! w^1 + (b+2s_2kv) \;  w^2\!\!\wedge\!w^3$, where $v=v(t)$ is the dynamical degree of freedom and $b$ is a constant. This field is homogeneous, axisymmetric and satisfy the identity $\mathrm{d}\mathbf{F}=\mathrm{d}^2\mathbf{A}=0$. Note that the field includes both an electric-type and a magnetic-type component which are parallel and pointing in the $w^1$ direction. Previous studies (which are restricted to the BI metric) have considered the case corresponding to $b=0$, ie. neglecting the possibility of a magnetic component.\footnote{Reference \cite{soda09mag} treated magnetic fields perturbatively.} Since our intention is to generalize the geometry, and not the matter fields, we will put $b=0$. In that case there is no magnetic-type component in the BI, BIII and KS spacetimes (since $s_2=0$). In the BII spacetime, however, where $s_2=1$, there is still a magnetic-type component. This is the minimal magnetic component required to satisfy the source-free Maxwell equations ($\mathrm{d}\mathbf{F}=0$). Our considered field strength is therefore:
\be
\mathbf F =  \dot v(t) \; w^0\!\!\wedge\! w^1 + 2s_2kv(t) \;  w^2\!\!\wedge\!w^3.
\ee
The corresponding vector potential in the gauge $A_0=0$ is $\mathbf{A} \equiv A_\mu w^\mu=v w^1$. 

We shall continue to specify all tensors, and perform all calculations, relative to the one-forms ($dt,w^1,w^2,w^3$).  In this basis the energy-momentum tensor is diagonal and we write:
\be
T^\mu_{\;\;\nu} = \text{diag}(-\rho, p_\parallel, p_\perp, p_\perp ).
\ee
Since $T^{\mu}_{\;\;\nu}$ is a diagonal mixed tensor of rank ($1,1$), the components are invariant under a normalization of the one-forms ($w^1,w^2,w^3$), ie. under a change to an orthonormal basis.  Thus the components of $T^{\mu}_{\;\;\nu}$ are physical quantities representing the energy density and pressure as measured in the fluid rest frame. We split the energy and pressure in the contributions from the scalar field ($\phi$) and vector field (A): 
\be
\begin{split}
\rho &= \rho_\phi + \rho_A, \\
p_\parallel &= p_\phi + (p_A)_\parallel, \\
p_\perp &= p_\phi + (p_A)_\perp.
\end{split}
\ee
The energy density and pressure of the scalar field takes the standard form in the entire family of spacetimes: 
\be
\rho_\phi = \frac{1}{2}\dot\phi^2 + V(\phi), \quad p_\phi=\frac{1}{2}\dot\phi^2 - V(\phi).  
\ee
The energy density (and pressure) of the vector field, on the other hand, depends on the spacetime:
\begin{equation} 
\rho_A = f^2e^{-2\alpha+4\sigma} \left( \frac{1}{2}\dot v^2  - s_2(\tr)v^2 \right).  
\label{rhoA}
\end{equation}
As $s_2\neq 0$ only for BII it turns out that the spatial curvature couples to the energy density only in this case. Clearly, this is due to the magnetic-type field unique for BII. Note that, since ${^3\!}R<0$ for BII, the coupling gives a positive contribution to the energy density.  The equation of state, however, is similar for all spacetimes: 
\begin{equation}
(p_A)_\parallel  = -\rho_A, \qquad (p_A)_\perp = + \rho_A. 
\label{eqofstateA}
\end{equation}
The relation $(p_A)_\parallel = -(p_A)_\perp$ also hold in the case of a \emph{massive} vector field \cite{dimo} (at least in the Bianchi type I metric).

The field equations (\ref{eqm1})-(\ref{eqm3}) for the considered family of spacetimes can be written:
\begin{table}[h]
\newcommand\T{\rule{0pt}{2.6ex}}
\newcommand\B{\rule[-1.2ex]{0pt}{0pt}}
\begin{center}
\begin{tabular}{ l c c}  
\hline\hline                       
Spacetime \T \B & $s_1$ & $s_2$ \\
  \hline
 BI \T & 0 & 0 \\
 BII  & 4 & 1 \\
 BIII \B $\&$ KS  & 1 & 0  \\
 \hline\hline
\end{tabular}
\caption{Spacetime dependent coefficients.}
\label{tabsi}
\end{center}
\end{table}
\begin{align}
&H^2 - \dot\sigma^2 =  \frac{\rho}{3M_p^2} - \frac{{^3\!}R}{6}, \label{hconstraint1} \\
&\dot H + 3H^2 =   \frac{1}{2M_p^2}(\rho - \frac{1}{3}p_\parallel - \frac{2}{3}p_\perp) -\frac{{^3\!}R}{3}, \label{doth} \\
&\ddot\sigma + 3H\dot\sigma =  \frac{p_\perp - p_\parallel}{3M_p^2} -  s_1 \frac{{^3\!}R}{6}, \label{shear} \\
&\dot\tr = -2(\dot\alpha+s_1\dot\sigma)\tr, \\
&\dot\rho_\phi + 3H(\rho_\phi+p_\phi) = Q \mathcal{L}_A \frac{2\dot\phi}{M_p}, \\
&\dot\rho_A + 4(H+\dot\sigma)\rho_A = -Q \mathcal{L}_A \frac{2\dot\phi}{M_p}, \label{s5}
\end{align}
where $\mathcal{L}_A=\frac{1}{2}f^2(\phi)e^{-2\alpha+4\sigma} \left( \dot v^2 + 2s_2^2 v^2 (\tr) \right)$.   From (\ref{shear}) it is manifest that the shear is sourced by anisotropic pressure \textit{and} anisotropic curvature.  
   
We will now introduce dimensionless variables and rewrite the field equations as an autonomous set of first order differential equations.  First we introduce the shear variable
\be
X\equiv\frac{H_\perp-H}{H} = \frac{\dot\sigma}{\dot \alpha},
\label{X}
\ee  
and a variable for the curvature:
\be
\Omega_K= -\frac{{^3\!}R}{6H^2}.
\label{K}
\ee
Note that $\Omega_K>0$ in all spacetimes apart from KS where it is negative (and BI, of course, where it is zero). Furthermore, we need the variables: 
\begin{equation}
Y=\frac{\dot \phi}{M_pH}, \quad Z=\frac{fe^{-\alpha+2\sigma} \dot v}{M_pH}, \quad \mathcal{V} = s_2\frac{ vfe^{-\alpha+2\sigma}}{M_p}.
\label{YZV}
\end{equation}
The variables X, Y and Z are similar to those used in \cite{soda10}.  The additional variables $\Omega_K$ and $\mathcal{V}$ are required to study the more general family of spacetimes considered here. We shall refer to the space spanned by the set of independent variables $(X,Y,Z,\Omega_K,\mathcal{V})$ as \emph{state space}. The space spanned by the constant parameters $\lambda$ and $Q$, we shall refer to as \emph{parameter space}.  Note that $\mathcal{V}\!\neq \! 0$ only in BII due to the coefficient $s_2$ in the definition of $\mathcal{V}$. It is only in BII, where ${^3\!}R$ couples to the energy density of the vector field, that the variable $\mathcal{V}$ is needed.

We can make use of the Hamiltonian constraint equation (\ref{hconstraint1}) to eliminate $V(\phi)$ from the equations of motions.  Using the identity $\frac{d\alpha}{dt} = H$ we change to the scale, $\alpha$, as time parameter, and the autonomous equations becomes: 
\begin{align}
&\frac{dX}{d\alpha} = \frac{1}{3} Z^2(X+1) + X \left[ 3(X^2-1) + \frac{1}{2} Y^2 \right]  + \Omega_K \left(s_1 + X + 4\mathcal{V}^2(1+X) \right), \label{seqX} \\
&\frac{dY}{d\alpha}= (Y+\lambda)\left(  3(X^2-1) + \frac{1}{2}Y^2  \right) + \frac{1}{3}YZ^2 + (Q+\frac{\lambda}{2})Z^2   \label{seqY} \\ 
&\qquad\quad +\Omega_K \left( Y + 3\lambda + 2\mathcal{V}^2\left( 3\lambda - 6Q + 2Y\right) \right), \notag \\
&\frac{dZ}{d\alpha}= Z\left[  3(X^2-1) + \frac{1}{2}Y^2 - Q Y + 1 - 2X + \frac{1}{3}Z^2 \right]  + \Omega_K \left[ Z + 4Z\mathcal{V}^2 - 12\mathcal{V} \right], \label{seqZ} \\
&\frac{d\Omega_K}{d\alpha} = 2\Omega_K\left[ -1 - s_1X + 3X^2 + \frac{1}{2} Y^2 + \frac{1}{3} Z^2  + \Omega_K + 4\Omega_K \mathcal{V}^2 \right], \label{seqN} \\
&\;\frac{d\mathcal{V}}{d\alpha} = (Q Y + 2X -1)\mathcal{V} + s_2 Z, \label{seqV} 
\end{align}
The dynamical variables are subject to the constraint:
\be
X^2 + \frac{1}{6}Y^2 + \frac{1}{6}Z^2 + \Omega_K \left(1+2 \mathcal{V}^2 \right) < 1,
\label{constraint}
\ee 
which follows from our considered case of a positive potential ($V(\phi)>0$). For BI, BII and BIII this implies absolute upper bounds on each of the variables since $\Omega_K \ge 0$. More specifically $X$ must be in the interval $(-1,1)$, Y and Z in $(-\sqrt{6}, \sqrt{6})$,  while $\Omega_K \left(1+2 \mathcal{V}^2 \right)$ in $(0,1)$.  In the KS spacetime, however, each of the variables might be arbitrarily large individually since $\Omega_K<0$.  Due to the negative curvature, the KS spacetime might collapse.  At the turning point, from expansion to contraction, we have $H\!\!=\!0$, and all the variables (apart from $\mathcal{V}$) diverges ($\rightarrow \infty$) as seen from the definitions (\ref{X})-(\ref{YZV}). For initial conditions leading to collapse, $\alpha$ is then usually not a suitable time-parameter. For our purposes, however, it is fine since we are just interested in \textit{whether}, and eventually \textit{when}, the universe collapse for a given set of initial conditions. 

It is only in BII that both ${^3\!R}\neq 0$ and $s_2\neq 0$. For the other spacetimes the equations simplifies somewhat:
\begin{align*} 
&\Omega_K \rightarrow 0  \;\quad \text{for BI},\\
&\mathcal{V} \rightarrow 0 \qquad  \text{for BI, BIII and KS}. 
\end{align*} 

Although we shall study the dynamics in terms of the independent variables ($X$, $Y$, $Z$, $\Omega_K$, $\mathcal{V}$), it is useful to introduce some auxiliary variables in order to make a closer connection to the physics. The (energy) density parameters  for the vector field and the scalar field can be expressed in terms of the independent variables in the following way:
\be
\Omega_A \equiv \frac{\rho_A}{3M_p^2H^2} = \frac{1}{6} Z^2 + 2\Omega_K\mathcal{V}^2 
\ee
and
\be
\Omega_\phi \equiv \frac{\rho_\phi}{3M_p^2H^2} = 1 - X^2 - \frac{1}{6} Z^2 -\Omega_K(1+2\mathcal{V}^2). 
\ee
Furthermore, we can split the latter one in the contributions from the kinetic and the potential energy: $\Omega_\phi = \Omega_{\text{kin}} + \Omega_V$, where $\Omega_{\text{kin}} = \frac{1}{6}Y^2$. The Hamiltonian constraint equation (\ref{hconstraint1}) can then be written on the generic form:
\be
1 = X^2 + \Omega_{\text{kin}} + \Omega_V + \Omega_A  + \Omega_K.
\label{hconstraint2}
\ee
It is also useful to express the deceleration parameter defined in (\ref{q}) in terms of the independent variables:
\be
q = -1 + 3X^2 + \frac{1}{2}Y^2 + \frac{1}{3}Z^2 + \Omega_K \left( 1 + 4\mathcal{V}^2 \right). 
\ee
From the definition of $\Omega_K$ it follows that
\be
\frac{d\Omega_K}{d\alpha} = -2\Omega_K(s_1 X - q).
\ee
We notice that, due to the shear , there is no guaranty for monotonically decaying curvature during inflation ($q\!<0$).  Nevertheless, as we shall see in the next section, the model turns out to be predictive since the potential energy of the scalar field $\Omega_V$ is monotonically increasing in a large region of state space. This leads the universe close to an anisotropic fix point with linear stability and vanishing curvature.

\section{Phase space analysis \label{chdsa1}}
Equipped with the field equations we shall now investigate the phase-space structure by dynamical system analysis and simulations. First we shall identify the fix points of the system and classify their linear stabilities. Although this is a powerful way to characterize phase space qualitatively, it gives unambiguous predictions only in the linear regime close to the fix points.  With arbitrary initial conditions it is therefore usually necessary to run simulations.  As we shall see, however, in our case the potential energy of the scalar field is monotonically increasing in a large region of state space for BI, BII and BIII, leading the system close to a stable and unique anisotropic fix point of type BI. The KS spacetime is a bit more complicated. Firstly, in this case $\Omega_V$ is not a monotone function. Secondly, the KS spacetime might collapse. We therefore investigate the phase flow in the KS universe by running simulations. Our results indicates that if the universe does not collapse, its fate is similar to that of the Bianchi type spacetimes. The model is therefore predictive and, as we shall demonstrate by simulations, the universe gets close to the stable anisotropic fix-point typically within a few e-folds.  

\subsection{Fix-points and linear stability\label{chdsa2}}
The fix points of the system are found by setting the left-hand side of the dynamical equations (\ref{seqX})-(\ref{seqV}) equal to zero and solving the algebraic equation. The stability is determined by linearizing the field equations around the fix-points, $\frac{d \delta X^i}{d\alpha}=\mathcal{M}\delta X^i$, and evaluating the eigenvalues of the matrix $\mathcal{M}$. If the real part of all eigenvalues are negative, the fix-point is stable and we call it an \emph{attractor}. If not all values are negative, the fix-point is unstable. Unstable fix-points are called \emph{saddles} if there are both positive and negative eigenvalues, and \emph{repellers} if all are positive. In general the phase flow goes from repellers, possibly via saddles, towards attractors. Without loss of generality we shall assume $\lambda>0$.\footnote{The situation $\lambda\rightarrow -\lambda$ is equivalent to $\phi\rightarrow -\phi$.} Under the assumption $\lambda\ll1$ we find $6$ fix-points satisfying the constraint (\ref{constraint}) or laying on its boundary. The fix-points are named (a)-(f) and the positions summarized in table \ref{tabfix1}. We shall first give a brief overview before studying each of them more carefully.

\begin{table}[h]
\newcommand\T{\rule{0pt}{2.6ex}}
\newcommand\B{\rule[-1.2ex]{0pt}{0pt}}
\begin{center}
\begin{tabular}{ l ccccc }
\hline \hline
Name \T \B &   $X$ & $Y$ & $Z^2$ & $\Omega_K$ & $\mathcal{V}$   \\ \hline
(a)\T  & $\sim\frac{\lambda Q-2}{3Q^2}$ & $\sim\frac{-2}{Q}$ & $\sim\frac{6\lambda Q -12}{2Q^2}$ & 0 & $\sim\frac{s_2Z}{3}$ \\
(b)     & $0$ & $-\lambda$ & $0$ & $0$ & $0$   \\
(c)     & $-1$ & $0$ & $\frac{18\lambda}{2Q-\lambda}$ & $-\frac{3(2Q+\lambda)}{2Q-\lambda}$ & $0$ \\
(d)     & $-\frac{2-\lambda^2}{2+2\lambda^2}$ & $-\frac{3\lambda}{1+\lambda^2}$ & $0$ & $-\frac{12-3\lambda^4}{4(1+\lambda^2)^2}$ & $0$ \\
(e)     & free & $\pm\sqrt{6-6X^2}$ & $0$ & $0$ & $0$ \\
(f)\B     & $\frac{1}{2}$ & $0$ & $0$ & $\frac{3}{4}$ & $0$  \\
\hline\hline
\end{tabular}
\caption{Coordinates of fix-points. For fix-point (a) we have written down an approximation. The exact position of (a) can be found in the text.}
\label{tabfix1}
\end{center}
\end{table}

\begin{table}[h]
\newcommand\T{\rule{0pt}{2.6ex}}
\newcommand\B{\rule[-1.2ex]{0pt}{0pt}}
\begin{center}
\begin{tabular}{ l ccccccccc }
\hline\hline 
Name \T \B &   Spacetime & Existence & Stability & $q$ & Comment  \\ \hline
(a)\T  & BI & $Q\gg1$ & attractor & $\sim(-1+\frac{\lambda}{Q})$  \\
(b)     &  FLRW  & - & - & $-1+\frac{\lambda^2}{2}$  \\
(c)     &  KS & $2Q>\lambda$ & saddle & $-1$  \\
(d)     &  KS & - & saddle & $-1+\frac{3\lambda^2}{2+2\lambda^2}$  \\
(e)     & BI & - & unstable & 2 & boundary \\
(f)\B     & BIII &-& saddle &$1/2$& boundary \\
\hline\hline
\end{tabular}
\caption{Properties of fix-points. The stability of (b) depends on $\lambda$ and $Q$ (it is a saddle when (a) exists, and stable if not).  See text for more details. }
\label{tabfix2}
\end{center}
\end{table}

\begin{figure*}[t]
\centering
\includegraphics[width=0.6\textwidth]{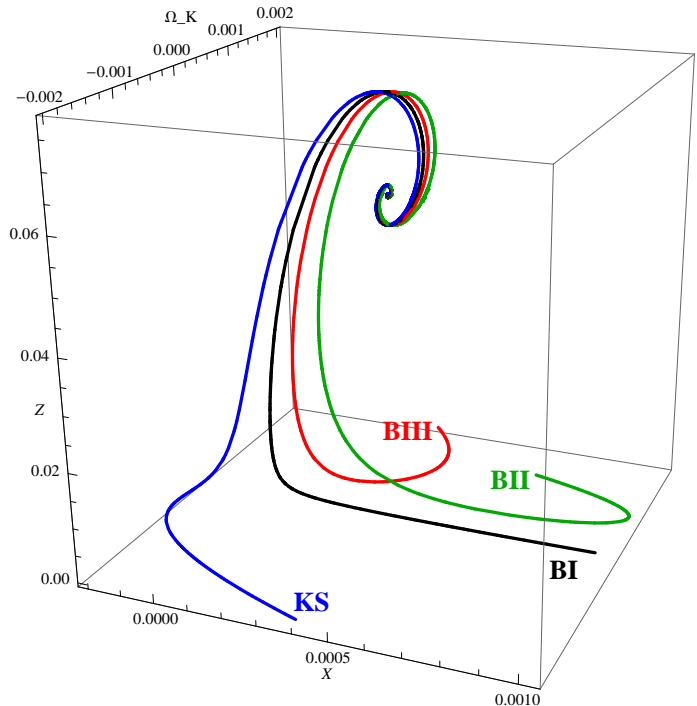}
\caption{The phase flow of $X$, $Z$ and $\Omega_K$ with $\lambda=0.1$ and $Q=50$.  The black, green, red and blue curves, respectively, represents simulations with BI, BII, BIII and KS initial conditions close to fixpoint (a). All trajectories converge towards a common point which is the type BI anisotropic fix point (a). }
\label{3dplot1}
\end{figure*}

\begin{figure*}[t]
\centering
\includegraphics[width=0.6\textwidth]{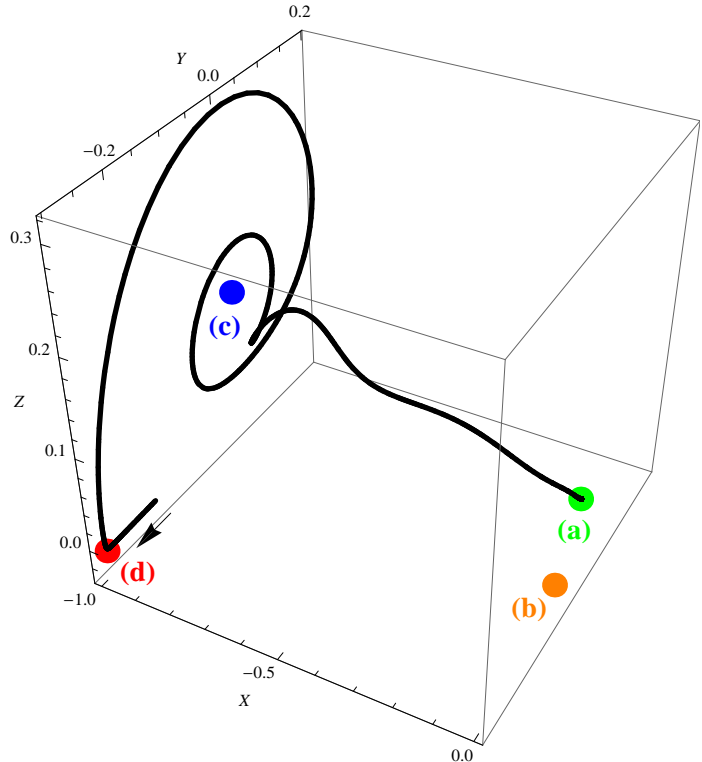}
\caption{The phase flow of $X$, $Y$ and $Z$ for KS initial conditions with $\lambda=0.1$ and $Q=50$.  Initial conditions and direction of flow is indicated by the arrow.  The fixpoint (a), (b), (c) and (d) are indicated by the colored points.  Initial conditions are carefully chosen such that the solution "rides" on both KS-saddles (d) and (c) before ending up in the BI anisotropic attractor (a).}
\label{3dplot2}
\end{figure*}

In table \ref{tabfix2} we give an overview of certain properties of the fix-points. Notice that the deceleration parameter $q\sim-1$ for (a)-(d) while $q>0$ for (e) and (f).  Thus (a)-(d) are inflationary fix-points, while (e) and (f) are decelerating. The fix-points (e) and (f) are on the boundary, ie. $V(\phi)=0$, while (a)-(d) satisfy the constraint (\ref{constraint}). Notice that there are no inflationary fix-points of type BII or BIII. For KS there are two inflationary fix-points, (c) and (d), but they are both saddles. A fix-point of special significance is the anisotropic attractor (a). Notice that it exists only when $Q\gg1$. Thus, interestingly, the shear $X$ is small if it exists. The only fine tuning in the model is therefore the usual $\lambda\ll 1$ required for slow-roll inflation.  No additional fine-tuning in any of the parameters is needed to avoid a dominating and observationally unacceptable shear. The model therefore represents a serious alternative to more conventional models. Fix-point (a) where identified in \cite{soda10} and the stability determined within the Bianchi type I framework. Note, however, that (a) is on the border to the entire family of spacetimes as all becomes type BI in the limit of zero curvature. It is therefore important to check the stability with respect to a broader class of spacetimes. As we shall see, (a) turns out to be the unique attractor for the entire family of spacetimes.  Loosely speaking, we can therefore say that the curved spacetimes becomes more and more Bianchi type I as they converges towards (a).  Technically, spacetimes of type BII, BIII or KS never becomes BI, of course, but with decaying curvature they can come arbitrarily close. See figure \ref{3dplot1} for a simulation close to the attractor (a).  Note how all spacetimes converges towards a common point which represents fix-point (a). In figure \ref{3dplot2} we show a simulation of type KS where the phase flow goes via the saddles (c) and (d) before converging towards (a). Fix-point (b) is isotropic and of type FLRW.  In \cite{soda10} it was shown that (b) is a saddle in the parameter region where (a) exists, and a stable attractor where (a) does not exist.  As we shall see this holds also in the more general class of spacetimes considered here (apart from BII if $Q\ll -1$).      

We shall now examine each of the fix-points more carefully.

\paragraph{Fixpoint (a)}
This is the anisotropically inflationary fix-point of Bianchi type I first identified in \cite{soda10}. In table \ref{tabfix1} the coordinates of (a) are given only to lowest order in the small quantities $\frac{\lambda}{Q}$ and $Q^{-2}$.  The exact position is:
\begin{align}
&X= \frac{2(\lambda^2 + 2Q\lambda -4)}{\lambda^2+8Q\lambda + 12 Q^2 +8},\\
&Y= -\frac{12(\lambda +2Q)}{\lambda^2 + 8Q\lambda +12Q^2 +8},\\
&Z^2= \frac{18(\lambda^2 + 2Q\lambda -4)(-\lambda^2 + 4Q\lambda + 12 Q^2 +8)}{(\lambda^2 + 8Q\lambda +12Q^2 +8)^2},\\
&\Omega_K= 0,\\
&\mathcal{V}= \frac{s_2Z}{1-2X-QY}.
\end{align}
Since the energy of the vector field is positive, ie. $\Omega_A>0$, we get the condition $Z^2>0$. The fix-point (a) therefore only exist in the parameter region where $2Q\lambda+\lambda^2>4$. Since $\lambda\ll1$ this implies $Q\gg 1$. As seen most directly from the approximations in table \ref{tabfix1}, it follows that (a) is near the origin of the state space ($X$,$Y$,$Z$,$\Omega_K$,$\mathcal{V}$). Consequently the fix-point is strongly dominated by the potential energy of the scalar field. For clarity we expand the eigenvalues in $\frac{\lambda}{Q}$ and $Q^{-2}$, and truncate at zero order. For BI the three eigenvalues are:
\be
\left( -3, \quad -\frac{3}{2}-i\sqrt{3(2\lambda Q +\lambda^2 -4) -\frac{9}{4}}, \quad -\frac{3}{2}+i\sqrt{3(2\lambda Q +\lambda^2 -4) -\frac{9}{4}}   \right).
\ee
For BII one has the additional eigenvalues $-3$ and $-2$, while for BIII and KS one has the additional eigenvalue $-2$.  The real part of all eigenvalues are negative for all spacetimes. Thus we have showed that the flat and anisotropic fix point  identified in \cite{soda10}, is an attractor also for the more general class of spacetimes considered here. Moreover, as we will verify below, it turns out to be the \textit{unique} attractor (when it exists) for our considered class.  Finally we mention that, as shown in \cite{soda10}, there is an exact power-law solution corresponding to this fix-point, where the line element takes the form:
\be 
ds^2 = -dt^2+t^{2k_1-4k_2}dx^2 + t^{2k_1+2k_2}(dy^2+dz^2),
\ee
where, to lowest order, $k_1\sim \frac{Q}{\lambda}$ and $k_2\sim\frac{\lambda Q -2}{3\lambda Q}$.

\paragraph{Fixpoint (b)}
This is a flat inflationary fix-point of type FLRW containing a scalar field dominated by its potential energy. The three eigenvalues for BI are:
\begin{equation*}
\left(  \frac{1}{2}(2\lambda Q + \lambda^2 -4), \quad -3+\frac{1}{2}\lambda^2, \quad -3+\frac{1}{2}\lambda^2  \right).
\end{equation*}
For BII one has the additional eigenvalues $-1-\lambda Q$ and $-2+\lambda^2$, while for BIII and KS one has the additional eigenvalue $-2+\lambda^2$.  This means that for the entire class of spacetimes, (b) is a saddle when (a) exists. If (a) does not exist, ie. $2\lambda Q + \lambda^2 -4 <0$, then (b) is an attractor for BI, BIII and KS. For BII it is also stable in a large parameter region if (a) does not exist, but not if $Q\ll-1$, in which case it is unstable.

\paragraph{Fixpoint (c)}
This is an inflationary fix-point of type KS containing an electric-type field and a cosmological constant (since the kinetic part of $\phi$ is vanishing). Essentially it is a generalization of the solutions with a pure cosmological constant found in \cite{linde88} and \cite{barrow96}. From the condition $Z^2>0$ it follows that the fix-point only exist in the parameter region $2Q>\lambda$. The four eigenvalues are:
\begin{equation*}
\left( -6,\quad 3,\quad -\frac{3}{2} + \frac{1}{2}\sqrt{1-8\lambda Q \frac{2Q+\lambda}{2Q-\lambda}}, \quad -\frac{3}{2} - \frac{1}{2}\sqrt{1-8\lambda Q \frac{2Q+\lambda}{2Q-\lambda}}   \right).
\end{equation*}
Notice that the real part of the two latter eigenvalues are always negative when $2Q>\lambda$, ie. when it exists.  We note that (c) is a saddle.

\paragraph{Fixpoint (d)}
This is an inflationary fix-point of type KS containing only our considered scalar field. First note that there are no condition on $Q$ for the existence of this fix-point. To lowest order in $\lambda$ the eigenvalues of (d) are:
\begin{equation*}
\Big(-3 +\mathcal{O}(\lambda^2) ,\quad -6 +\mathcal{O}(\lambda^2),\quad 3 +\mathcal{O}(\lambda^2),\quad 3\lambda Q +\mathcal{O}(\lambda^2)    \Big).  
\end{equation*}
Since $\lambda\ll 1$ the first and second eigenvalues are negative, while the third are positive.  The sign of the last eigenvalue depends on $Q$. In any case (d) is a saddle. Although our analysis focus on slow-roll inflation $\lambda\ll1$, we mention that there exist a region in parameter space (where $\lambda$ is larger than unity) where all the eigenvalues are negative and the fix point is a stable attractor. In this region, however, the deceleration parameter is positive and the fix-point is not inflationary. 

\paragraph{Fixpoint (e)} 
This is a decelerating fix-point of type BI containing only a pure kinetic scalar field (thus the fluid is stiff $\rho=p$). The solution is part of a broader solution commonly referred to as \emph{Jacobs disc} \cite{jacobs68}.  See also \cite{hervik01} for a discussion of such solutions.  Note that (e) is a \textit{curve} of fix-points, satisfying $6X^2 + Y^2=6$. Since $V(\phi)=0$, on the curve, it lays on the boundary of our considered state space.  We have two sets of eigenvalues depending on the sign of $Y=\pm\sqrt{6-6X^2}$. The eigenvalues for BI:
\begin{equation*}
\left(  0, \quad 1-2X \mp Q\sqrt{6-6X^2}, \quad 6\pm\lambda\sqrt{6-6X^2}  \right)
\end{equation*}
The sign of the third eigenvalue is always positive since $\lambda\ll1$, while the second eigenvalue depends on $Q$ and the sign of $Y$.  For BII one has the additional eigenvalues $4-8X$ and $-1+2X\pm Q\sqrt{6-6X^2}$, while for BIII and KS one has the additional eigenvalue $4-2s_1 X$.  In any case (e) is unstable. 

\paragraph{Fixpoint (f)} 
This is a decelerating fix-point of type BIII containing no fluids. Essentially it is a Bianchi type III generalization of the Milne universe (but with a trivial flat direction).  Like (e), it lays on the boundary $V(\phi)=0$. The four eigenvalues are:
\begin{equation*}
\left(   3, \quad -3/2, \quad -3/2, \quad -3/2   \right).
\end{equation*}
Thus (f) is a saddle.

\subsection{Phase flow with arbitrary initial conditions \label{chdsa3}}

\begin{table}[H]
\footnotesize
\newcommand\T{\rule{0pt}{2.6ex}}
\newcommand\B{\rule[-1.2ex]{0pt}{0pt}}
\begin{center}
\begin{tabular}{l c c c c c c c c}
\hline\hline
\T \B &  \multicolumn{5}{c}{Initial conditions} & & \multicolumn{2}{c}{$D<\frac{\lambda}{Q}$ at time ($\alpha$)}    \\ \cline{2-6} \cline{8-9}
     Spacetime \T \B    & $X$ & $Y$ & $Z$ & $\Omega_K$ & $\mathcal{V}$ &  & $Q=50$ & $Q=1000$ \\ \hline
BI \T & $\;\;\;0.4$ & $-0.6$ & $\;\;\;0.5$ & - &  - & & $\;\;6.0$  & $5.7$\\
BI & $-0.4$ & $-1.1$ & $\;\;\;1.5$ & - &  - & & $13.4$  & $6.0$\\
BI & $-0.4$ & $\;\;\;0.3$ & $-0.7$ & - &  - & & $\;\;6.1$  & $5.8$\\
BI & $\;\;\;0.1$ & $\;\;\;0.7$ & $-0.7$ & - &  - & & $\;\;7.4$  & $5.8$\\
BI & $\;\;\;0.3$ & $-1.4$ & $-0.1$ & - &  - & & $\;\;9.5$  & $5.8$\\
BII & $-0.2$ & $-0.3$ & $\;\;\;0.5$ & $0.1$ & $-0.1$ &  & $\;\;2.9$  & $4.7$\\
BII & $-0.3$ & $-1.1$ & $-1.2$ & $0.2$ & $-0.4$  & & $\;\;3.7$  & $5.2$\\
BII & $-0.1$ & $\;\;\;0.1$ & $-0.9$ & $0.1$ & $-0.1$ & & $\;\;3.3$  & $4.8$ \\
BII & $-0.5$ & $\;\;\;0.2$ & $-1.1$ & $0.1$ & $-0.1$ & & $\;\;3.5$  & $5.2$\\
BII & $-0.6$ & $\;\;\;0.8$ & $-0.5$ & $0.3$ & $-0.2$ & & $\;\;3.8$  & $5.0$\\
BIII & $-0.5$   & $-0.8$ & $\;\;\;0.6$ & $0.1$ & - & & $\;\;8.1$ & $5.8$\\
BIII & $\;\;\;0.1$ & $\;\;\;0.8$ & $-0.4$ & $0.2$ & - & & $\;\;7.4$ & $5.7$\\
BIII & $-0.1$ & $-1.5$ & $-1.3$ & $0.2$ & - & & $12.7$ & $6.5$\\
BIII & $-0.6$ & $-1.7$ & $-0.3$ & $0.1$ & - & & $14.1$ & $6.6$\\
BIII \B & $\;\;\;0.3$ & $-0.9$ & $\;\;\;0.5$ & $0.3$ & - & & $\;\;8.2$ & $5.9$\\
\hline\hline
\end{tabular}
\caption{Simulations for Bianchi type spacetimes with initial conditions generated by a random number generator. $D$ is the distance to fix-point (a) defined in (\ref{distance}). When $D<\frac{\lambda}{Q}$ the universe is in the linear regime close to fix-point (a), and the distance to (a) is much smaller than the distance to (b). Two runs with different values for $Q$, while $\lambda=0.1$ in all runs. }
\label{tabsim1}
\end{center}
\end{table}

\begin{table}[p]
\footnotesize
\newcommand\T{\rule{0pt}{2.6ex}}
\newcommand\B{\rule[-1.2ex]{0pt}{0pt}}
\begin{center}
\begin{tabular}{ r c c  c  c  c c c  c c c  c }
\hline\hline
\T \B &&  \multicolumn{4}{c}{Initial conditions} && \multicolumn{2}{c}{$D<\frac{\lambda}{Q}$ at time ($\alpha$)} & & \multicolumn{2}{c}{$H=0$ at time ($\alpha$)}   \\ \cline{3-6} \cline{8-9}\cline{11-12} 
\# \T \B &   &  $X$ & $Y$ & $Z$ & $\Omega_K$ & & $Q\!=\!50$ & $Q\!=\!1000$  & & $Q\!=\!50$ & $Q\!=\!1000$   \\ \hline
\T1&&$\;\;\;0.6$  & $-1.5$ & $\;\;\;2.2$ & $-0.8$ && $12.7$ & $6.4$ && $-$ & $-$\\
2&&$\;\;\;0.8$  & $\;\;\;1.5$  & $-0.4$ & $-0.7$&& $12.4$ & $5.8$ && $-$ & $-$\\
3&&$\;\;\;0.6$  & $-1.4$ & $\;\;\;0.3$ & $-0.5$ && $\;\;9.4$ & $6.0$ && $-$ & $-$\\
4&&$-0.1$ & $\;\;\;2.1$  & $\;\;\;2.4$ & $-0.9$ && $-$ & $-$ && $0.3$ & $0.3$\\
5&&$\;\;\;0.7$  & $-0.5$ & $\;\;\;2.0$ & $-0.9$ && $12.7$ & $6.1$ && $-$ & $-$\\ 
6&&$\;\;\;2.2$ & $\;\;\;4.0$ & $\;\;\;3.9$ & $-9.3$ && $-$ & $-$ && $0.03$ & $0.03$ \\
7&&$\;\;\;1.6$ & $-2.5$ & $-0.2$ & $-3.1$ && $-$ & $-$ && $0.13$ & $0.13$ \\
8&&$\;\;\;1.5$ & $-4.9$ & $-3.8$ & $-8.7$ && $-$ & $-$ && $0.03$ & $0.04$ \\
9&&$-0.8$ & $-1.8$ & $\;\;\;2.6$ & $-2.3$ && $-$ & $-$ && $0.17$ & $0.17$ \\
10&&$-0.9$ & $\;\;\;2.7$ & $-4.9$ & $-5.1$&& $-$ & $-$ && $0.04$ & $0.04$ \\ 
11 &&  $\;\;\;0.6$ & $-1.4$ & $\;\;\;0.7$ & $-23$ && $3.8$ & $5.5$ && $-$ & $-$ \\
12 &&  $\;\;\;0.2$ & $-0.6$ & $-1.7$ & $-22$ && $4.1$ & $5.6$ && $-$ & $-$ \\
13 &&  $\;\;\;0.7$ & $\;\;\;2.0$ & $\;\;\;1.9$ & $-28$ && $5.2$ & $5.9$ && $-$ & $-$ \\
14 &&  $\;\;\;0.9$ & $\;\;\;1.9$ & $-0.1$ & $-21$ && $5.1$ & $5.5$ && $-$ & $-$ \\
15\B &&  $-0.7$ & $-2.2$ & $-1.0$ & $-24$ && $4.5$ & $6.0$ && $-$ & $-$ \\
\hline\hline
\end{tabular}
\caption{Simulations for KS with various initial conditions. $D$ is the distance to fix-point (a) defined in (\ref{distance}). Two runs with different values for $Q$, while $\lambda=0.1$ in all runs. Initial values for $\Omega_K$ where picked uniformly by a random number generator from the intervals $(-1.0,\; -0.1)$, $(-10.0,\; -2.0)$ and $(-30,\; -20)$, respectively, for simulations $\#$1-5, $\#$5-10 and $\#$11-15.  In simulations $\#$6-10 the potential energy of the scalar field is initially sub-dominant, while in $\#$11-15 $\Omega_V\sim |\Omega_K|$. } 
\label{tabsim2}
\end{center}
\end{table}

The stability analysis above determines the phase flow close to the fix-points. With arbitrary initial conditions, however, we need something more to determine the fate of the dynamical system.  For the spacetimes of type BI, BII and BIII we shall now see that far away from (a) the phase flow can be characterized by the potential energy of scalar field. 

In section \ref{cheq} we defined an auxiliary variable $\Omega_V$, representing the potential energy of the scalar field. The equation of motion is:
\be
\frac{d\Omega_V}{d\alpha} = \Omega_V \left[ \lambda Y + 2q + 2 \right] = \Omega_V F, 
\label{auxiliary}
\ee
where $F= 6X^2 + (Y+\frac{\lambda}{2})^2 - \frac{\lambda^2}{4} + \frac{2}{3}Z^2 + 2\Omega_K(1+4\mathcal{V}^2)$. Note that if $\Omega_K\!\ge\!0$ \textit{and} $|X^i|\gg\lambda$ for at least one of the variables, then $F>0$, and consequently, $\Omega_V$ is monotonically increasing. Note that, as implied by (\ref{hconstraint2}), $\Omega_V\!\sim \!1$ when all the independent variables are small ($\ll 1$).  Regardless of initial conditions, the system will therefore approach some solution where $\Omega_V \sim 1$ and all the independent variables are small ($\ll \!1$). The monotone function $\Omega_V$ will therefore lead the system close to the origin of the state space ($X$,$Y$,$Z$,$\Omega_K$,$\mathcal{V}$). In this region we have the stable anisotropic fix-point (a). The system will therefore eventually end up at the anisotropic fix-point (a) for arbitrary, but non-special\footnote{One can construct special initial-conditions not leading to (a). With initially vanishing vector field for instance, $Z=0$, the universe will converge towards the isotropic fix-point (b) instead of (a).}, initial conditions.

In principle, it could take arbitrary long time for the system to get close to (a), since it might spend long time at the some of the saddles (for example at (b) which is also located close to the origin). Simulations, however, demonstrates that the typical time spent to get close to (a), starting with arbitrary initial conditions, only represents a minor fraction of the total number of e-folds during inflation. To demonstrate this, we define the distance to the fix-point (a) by
\be
D=\sqrt{(X-X_a)^2+(Y-Y_a)^2+(Z-Z_a)^2+(\Omega_K-(\Omega_K)_a)^2+(\mathcal{V}-\mathcal{V}_a)^2},
\label{distance}
\ee
where $X_a$ is the value of $X$ at (a) and similar for the other coordinates. Note that the distance between (a) and (b) is of order $\sim\lambda \sim Q^{-1}$.  We say that the system is close to (a) when $D<\frac{\lambda}{Q}$. With this definition we ensure that the system is effectively unaffected by (b) and attracted by (a). In table \ref{tabsim1} we have summarized some simulations with $\lambda=0.1$ and  two different values for $Q$. For $Q=50$ we note that it typically takes 4-14 e-folds for the system to get close to fix-point (a).  For $Q=1000$ it typically takes 5-7 e-folds.  

The above analysis does not apply for KS since $\Omega_V$ is not a monotone function when $\Omega_K<0$. Instead we study the KS spacetime by numerical simulations.  Some results for randomly chosen initial conditions are summarized in table \ref{tabsim2}.  As mentioned above, the KS universe might collapse, in which case $H\rightarrow 0$. Simulations shows that non-special initial conditions, either lead to collapse or the system will converge towards the fix-point (a). For initial conditions leading to collapse, the table shows how long time it takes before $H=0$. We note that $H\rightarrow 0$ very quickly, typically within the first third of the first e-fold, if the universe collapses.  For initial conditions not leading to collapse, simulations shows that the system ends up at (a), and the table shows how long time it takes to get a distance $D<\frac{\lambda}{Q}$ from (a).  The results are very similar as those for BI, BII and BIII if the universe does not collapse. The system is close to (a) typically after 4-14 e-folds for $Q=50$ and 5-7 e-folds for $Q=1000$.

Supported by simulations we have showed that for initial conditions leading to interesting cosmologies (no immediate recollapse), the universe gets close to (a) relatively quickly.\footnote{We should mention that for several of the (randomly chosen) initial conditions in tables \ref{tabsim1} and \ref{tabsim2}, the universe is not accelerating, ie. $q$ might be positive at the initial time $\alpha=0$. The time taken to get close to (a) after the start of inflation, is therefore somewhat smaller than the times in the table for the cases where $q>0$ initially. In any case, since we typically assume inflation lasted for at least $60$ e-folds, only a minor fraction will be spend far away from (a).} When the system is close to (a) we see from (\ref{seqN}) that the curvature decays exponentially, $\Omega_K\sim e^{-2\alpha}$. If inflation lasted for $N$ e-folds, a rough estimate for the curvature at the end of inflation is therefore $(\Omega_K)_{\text{end}} \sim e^{-2N}$.

\section{Late time behavior \label{chlate}}
To check if the model gives rise to an acceptable cosmology we will now investigate the late-time behavior.  We showed above that at the end of inflation the universe will be extremely close to the anisotropic fix-point (a), with a curvature of order $(\Omega_K)_{\text{end}} \sim e^{-2N}$ if inflation lasted for N e-folds.  At the end of inflation we assume a period of reheating where the energy of the scalar and vector fields are dumped primarily into radiation. Reheating is a poorly understood process, but we shall assume that it is sufficiently fast that we can take the initial conditions for the shear ($X$) and the curvature ($\Omega_K$) to be similar as at the end of inflation. In our analysis we shall, for simplicity, introduce a single perfect fluid with equation of state $p=\omega\rho$. The radiation dominated era after inflation can then be studied by setting $\omega=\frac{1}{3}$, while the subsequent period dominated by non-relativistic fluids can be studied by setting $\omega=0$. We shall neglect the cosmological constant, since it would not change the estimates we seek significantly.  

The conservation equation in the class of spacetimes yields
\be
\dot\rho +3H(1+\omega)\rho = 0.
\ee
The associated density parameter is defined
\be
\Omega_M = \frac{\rho}{3M_p^2H^2}.
\ee 
The Hamiltonian constraint equation can then be written:
\be
1 = X^2 + \Omega_M + \Omega_K,
\label{late1}
\ee
where $X$ and $\Omega_K$ is defined in the same way as above. Due to the constraint (\ref{late1}) the system is effectively two dimensional, and we choose $\Omega_M$ as the auxiliary variable. The autonomous equations for the post-inflationary era with a single perfect fluid can then be written: 
\begin{align}
&\frac{dX}{d\alpha} =  - \frac{1}{2}X\left[ (1+3\omega) \Omega_K  +3(1-\omega)(1-X^2) \right] + s_1 \Omega_K ,\label{late2}\\ 
&\frac{d\Omega_K}{d\alpha} = \Omega_K \left[  (1-\Omega_K)(1+3\omega) -2s_1X +3(1-\omega)X^2 \right]. \label{late3}
\end{align}

\begin{table}[h]
\newcommand\T{\rule{0pt}{2.6ex}}
\newcommand\B{\rule[-1.2ex]{0pt}{0pt}}
\begin{center}
\begin{tabular}{l l l l l l}
\hline\hline
Name \T \B & Spacetime    & $X$                               &    $\Omega_K$                         &    $\Omega_M$              &    Stability   \\ \hline
(P1) \T         & FLRW            &  $0$                              &    $0$                                          &    $1$                              &    saddle    \\
(P2) \T         & BII                   &  $\frac{1}{8}(1\!+\!3\omega)$  &    $\frac{3}{64}(1\!+\!2\omega\!-\!3\omega^2)$  &    $\frac{15-3\omega}{16}$   &    attractor   \\
(P3) \T         & BIII                  &  $\frac{1}{2}$               &    $\frac{3}{4}$                          &    $0$                              &    attractor \\
(P4) \T         & BI                    &  $\pm 1$                      &     $0$                                         &    $0$                              &    unstable \\
\hline\hline
\end{tabular}
\caption{Post-inflationary fix-points. The unstable fix-point (P4) is either a saddle or a repeller depending on the spacetime and the sign of $X$ (see eigenvalues). }
\label{tabpostinflation}
\end{center}
\end{table}

\begin{table}[h]
\newcommand\T{\rule{0pt}{2.6ex}}
\newcommand\B{\rule[-1.2ex]{0pt}{0pt}}
\begin{center}
\begin{tabular}{l l l}
\hline\hline
Name \T \B & Spacetime    & Eigenvalues                                                                                 \\ \hline
(P1) \T         & FLRW            &  $\left(  1+3\omega, \quad -\frac{3}{2}(1-\omega)  \right)$   \\
(P2) \T         & BII                   &  $\left(  -\frac{3}{4}(1-\omega), \quad -\frac{3}{4}(1-\omega)  \right)$   \\
(P3) \T         & BIII                  &  $\left(  -\frac{3}{2}, \quad -3\omega  \right)$               \\
(P4) \T         & BI                    &  $\left( 4\mp 2s_1, \quad 3(1-\omega)  \right)$             \\
\hline\hline
\end{tabular}
\caption{Real part of eigenvalues for post-inflationary fix-points. }
\label{tabpostinflation2}
\end{center}
\end{table}

This system has four fix-points (P1)-(P4) summarized in table \ref{tabpostinflation}. The real parts of the eigenvalues can be found in table \ref{tabpostinflation2}. Note that there are no fix-points of type KS. Without introducing a dark energy in the model, the KS spacetime will eventually collapse \cite{collins77}. The Bianchi type I fix-point (P4) represents the two special points on the Kasner circle which are LRS \cite{ellisboka}. Also note that it is a special case of (e) in table \ref{tabfix1}. Setting $\omega=0$, the attractors (P2) and (P3) represents the late time solutions for BII and BIII, respectively, for a universe without dark energy. Fix-point (P2) corresponds to the Collins-Stewart Bianchi type II exact solution \cite{collins71}, while (P3) is similar as fix-point (f) in table \ref{tabfix1}. For $\omega=0$, (P2) and (P3) are the global attractors for the most general perfect fluids of type II and III, respectively, even including tilt \cite{tilt1}, \cite{tilt2}. 

\begin{figure*}[t]
\centering
\includegraphics[width=0.6\textwidth]{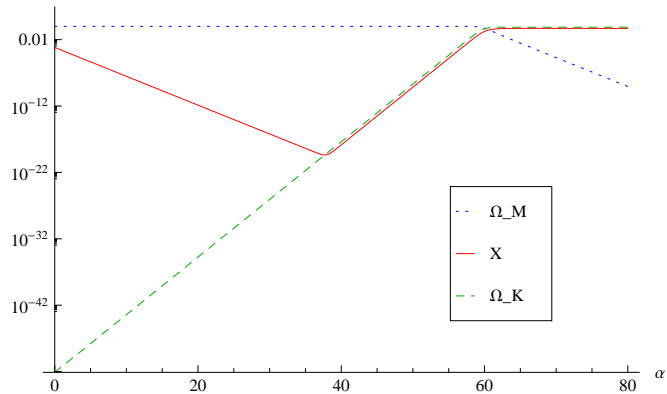}
\caption{Simulation of BIII after inflation with equation of state $\omega=1/3$ for the perfect fluid.  The initial conditions comes from an inflation model with $\lambda=0.1$ $Q=50$ and number of e-folds $N=60$.}
\label{lateb3}
\end{figure*}

Note that the shear in the attractors (P2) and (P3), with $X$ around unity, is quite extreme.  Supernova Ia data gives the bound $-0.012<X_\text{today}<0.012$ to one sigma confidence level for a Bianchi type I spacetime with rotational symmetry \cite{supernova10}. The goal of this section is to find the minimum number of e-folds during inflation, $N_\text{min}$, required for a shear today in agreement with supernova observations. For this estimate we shall use $|X_\text{today}|<0.01$. More stringent bounds can probably be found from the CMB which is very sensitive to shear \cite{barrow85,mota08,mota08nr2,koivi1}. For simplicity we shall use the supernovae since the final result is not very sensitive to the bound on $X_\text{today}$.  Thus, observational bounds certainly implies that the universe must have been very close to the flat FLRW saddle (P1) just before dark energy became significant.  To determine $N_\text{min}$ we must therefore investigate the dynamics close to the fix-point (P1).  Linearizing (\ref{late2})-(\ref{late3}) around (P1) yields:
\begin{align}
&\frac{dX}{d\alpha} = -\frac{3}{2}(1-\omega)X + s_1  \Omega_K,  \label{late22}\\
&\frac{d \Omega_K}{d\alpha} = (1+3\omega)  \Omega_K. \label{late33}
\end{align}
These equations can be solved exactly:
\begin{align}
& X(\alpha) =  e^{-\frac{3}{2}(1-\omega)\alpha}\left[  \pm \frac{2s_1}{5+3\omega} e^{\frac{1}{2}(5+3\omega)\alpha-2N}(1-e^{-\frac{1}{2}(5+3\omega)\alpha}) + \frac{\lambda Q-2}{3Q^2}  \right], \label{exact1} \\
& \Omega_K(\alpha) = \pm e^{(1+3\omega)\alpha - 2N}, \label{exact2}
\end{align}
where we have used the initial conditions $X(0)= \frac{\lambda
  Q-2}{3Q^2}$ and $\Omega_K(0) = \pm e^{-2N}$. Here $N$ is the number
of e-folds during inflation.  The initial condition for the shear
corresponds to the shear at the anisotropic fix-point (a), while for
the curvature we have used the estimate at the end of inflation
derived above. To get some intuition for the solution we shall now set
$\omega = 1/3$ and study the behavior at different times. In the
regime $0\!<\!\alpha\!<\!\frac{2}{3}N$ we have the approximation
$X\sim \frac{\lambda Q-2}{3Q^2} e^{-\alpha}$.  Thus there is an era of
isotropization starting right after reheating, where the shear decays
as $\sim e^{-\alpha}$, lasting until $\alpha\sim \frac{2}{3} N$. Also
notice that $X\sim \Omega_K$ around $\alpha=\frac{2}{3}N$. This is
also true for KS since the sign of $X$ will change from positive to
negative around the same time. In the period
$\frac{2}{3}N\!<\!\alpha\!<\!N$ we have the approximation $X \sim \pm
\frac{1}{3}s_1 e^{2\alpha-2N}$.  In this period we have $\ln{|X|} \sim
\ln{|\Omega_K|} \sim 2\alpha-2N$.  We therefore say that the curvature
"tracks" the shear in this period.  Around $\alpha=N$ the parameters
$X$ and $\Omega_K$ will approach unity and the universe is no longer
close to the matter dominated fix-point (P1). The solutions
(\ref{exact1})-(\ref{exact2}) are then no longer valid. In the regime
$\alpha>N$ the universe will in general converge towards the
attractors (P2) and (P3), respectively, for BII and BIII, while KS
will eventually collapse.  Simulations of the autonomous equations
(\ref{late2})-(\ref{late3}) verifies these approximations. For an
example with BIII spacetime, see figure (\ref{lateb3}). Note how the
shear decays until it catches up with the curvature slightly before
$\alpha=\frac{2}{3}N$. After that the shear is tracked by the
curvature, approaching unity around $\alpha\sim N$.  This behavior is
in close agreement with the above approximations.  It is easily seen
from the solutions (\ref{exact1})-(\ref{exact2}) that the same
tracking behavior occurs also with $\omega \neq 1/3$. Thus the shear
will be tracked by the curvature also in the matter dominated era.

Reheating is complete when practically all of the energy is in radiation at thermal equilibrium. We shall denote the time and temperature at this stage by $\alpha_\text{reh}$ and $T_\text{reh}$, the reheat time and temperature, respectively.  Furthermore we shall denote the time at radiation-matter equivalence by $\alpha_\text{eq}$, and today by $\alpha_0$.  Setting the initial time $\alpha_\text{reh}=0$ at $T=T_{\text{reh}}$ we get $\alpha_0 = \ln{\frac{T_\text{reh}}{T_0}}$, where $T_0=2.73 K$ is the temperature of the CMB today. Here we have used that the temperature of a photon gas scales as $\propto e^{-\alpha}$. Matter-radiation equivalence occurred around redshift $3300$ corresponding to  $\alpha_\text{eq} \sim \alpha_0 -8$.  We can then find the minimum number of e-folds during inflation, $N_\text{min}$, required for a shear today consistent with supernova data, ie. $|X_0|<10^{-2}$. The initial condition for the curvature is $\Omega_K(\alpha_\text{reh}) = e^{-2N}$. Assuming that the universe is close to the saddle (P1) all the way up to today, the curvature grows approximately as $\Omega_K \propto e^{2\alpha}$ in the radiation dominated era, and as $\Omega_K \propto e^{\alpha}$ in the matter dominated era.  The curvature today is then $(\Omega_K)_0 \sim e^{\alpha_\text{eq} + \alpha_0 -2N}$. Since the shear is tracked by the curvature both in the radiation and matter dominated eras, we have $X_0 \sim (\Omega_K)_0$. Let $|X_\text{max}|$ denote the maximum shear today consistent with observations. The condition $|X_0|\le|X_\text{max}|$, then leads to: 
\be
N_\text{min} = \ln{\frac{T_\text{reh}}{T_0}}-\frac{1}{2}\ln |X_\text{max}| - 4.
\ee
As discussed above we shall use $|X_\text{max}|=10^{-2}$, consistent with the supernova Ia data. A lower bound on the reheat temperature $T_\text{reh}\gtrsim 1\text{MeV}$ comes from the requirement that reheating must have occurred before the start of nucleosynthesis. This corresponds to $N_\text{min}=21$.  There is also an upper limit $T_\text{reh}< 10^9\text{GeV}$ if supersymmetry exists \cite{boka}, corresponding to $N_\text{min}=48$.  

To summarize, we have showed that plausible bounds on $T_\text{reh}$ implies $N_\text{min}$ in the interval $(21,48)$.

\section{Conclusion}

In this paper we have studied an inflationary scenario with a stable anisotropic hair.  Since the model provides a counter-example to the cosmic no-hair theorem, it is important to study the model with more general initial conditions. Ultimately, as suggested in \cite{soda09,soda10}, this might lead to a modified cosmic no-hair theorem.  As a first step we took a dynamical system approach and showed that the stable anistropic fix-point of type BI, identified in \cite{soda10}, is the unique attractor for a wider class of spacetimes exhibiting spatial curvature.  Moreover, for the considered Bianchi type spacetimes, we showed that the potential energy of the scalar field ($\Omega_V$) is monotonically growing in a large variable region leading the universe close to the origin of the state space. In this region we have the stable anisotropic fix-point. This observation explains the rapid convergence in numerical simulations with arbitrary initial conditions, and, in fact, it is a reminiscent of the cosmic no-hair theorem itself.\footnote{The only difference is that, while the no-hair theorem relies only on the fact that $\Omega_V$ is a monotone function, $\Omega_V$ is only monotone until it is almost $1$ in our considered model. Our argument therefore also relies on the dynamical system analysis.} The KS spacetime, however,  which (depending on initial conditions) might collapse, is more complicated and our analysis relies on numerical simulations. For initial conditions not leading to collapse, our simulations shows that the behavior is quite similar as for the considered Bianchi type spacetimes. For arbitrary, but non-special\footnote{For example, if the energy density of the vector field is initially vanishing, it will always be vanishing, and the universe goes to the isotropic FLRW saddle point.}, initial conditions, the universe will typically get close to the attractor after a few e-folds for any of the considered spacetimes. Thus,  for the major fraction of e-folds required for inflation, the inflationary universe is well described by this fix-point. 

Thanks to these features the model provides unambiguous initial conditions for the era after inflation. After reheating, when the energy of the scalar and vector fields are dumped primarily into radiation, the only hair left is the anisotropic expansion. Interestingly, these initial conditions yields an acceptable late-time cosmology. After reheating there will be a period of isotropization lasting for $\sim\frac{2}{3}N$ e-folds, where $N$ is the number of e-folds during inflation. After that the shear scales as the curvature and becomes dominant around $N$ e-folds after the end of inflation. For plausible bounds on the reheat temperature, the minimum number of e-folds during inflation, required for consistency with the isotropy of the supernova Ia data, lays in the interval ($21,48$). As a rule of thumb, successful inflation models must be stable for at least 60 e-folds.  It is clear that this number is sufficient also for the considered model despite the stable anisotropic hair.  So, to answer the question in the title: \emph{yes}, the results obtained for our restricted class of spacetimes indicates that inflation with anisotropic hair is cosmologically viable.

\acknowledgments{
DFM thanks the Research Council of Norway FRINAT grant 197251/V30. DFM is also partially supported by project PTDC/FIS/111725/2009 and CERN/FP/116398/2010. We also appreciate useful comments from an anonymous referee.
}


\begin{thebibliography}{99}\footnotesize


\bibitem{Efstathiou04}
G. Efstathiou,
  \emph{A Maximum Likelihood Analysis of the Low CMB Multipoles from WMAP},
   MNRAS \textbf{348} (2004) 885 [\href{http://arxiv.org/abs/astro-ph/0310207}{arXiv:astro-ph/0310207}]. 

\bibitem{oliveira04}
A. de Oliveira-Costa, M. Tegmark, M. Zaldarriaga and A. Hamilton,
  \emph{The significance of the largest scale CMB fluctuations in WMAP},
   Phys. Rev. D \textbf{69} (2004) 063516  [\href{http://arxiv.org/abs/astro-ph/0307282}{arXiv:astro-ph/0307282}].

\bibitem{copi04}
C. J. Copi, D. Huterer and G. D. Starkman, 
\emph{Multipole Vectors--a new representation of the CMB sky and evidence for statistical anisotropy or non-Gaussianity at $2<=l<=8$},
Phys. Rev. D \textbf{70} (2004) 043515 [\href{http://arxiv.org/abs/astro-ph/0310511}{arXiv:astro-ph/0310511}].

\bibitem{eriksen04}
H. K. Eriksen, F. K. Hansen, A. J. Banday, K. M. Gorski and P. B. Lilje,
  \emph{Asymmetries in the CMB anisotropy field},
  Astrophys. J. \textbf{605} (2004) 14,
    [Erratum-ibid. \textbf{609} (2004) 1198] [\href{http://arxiv.org/abs/astro-ph/0307507}{arXiv:astro-ph/0307507}].

\bibitem{hansen04}
F. K. Hansen, A. J. Banday and K. M. Gorski,
  \emph{Testing the cosmological principle of isotropy: local power spectrum estimates of the WMAP data},
MNRAS \textbf{354} (2004) 641 [\href{http://arxiv.org/abs/astro-ph/0404206}{arXiv:astro-ph/0404206}].

\bibitem{nico1}
  N.~E.~Groeneboom, M.~Axelsson, D.~F.~Mota and T.~Koivisto,
  \emph{Imprints of a hemispherical power asymmetry in the seven-year WMAP data due to non-commutativity of space-time},
    \href{http://arxiv.org/abs/1011.5353}{arXiv:astro-ph/1011.5353}.
    
\bibitem{dav4}
  L.~H.~Ford,
  \emph{Inflation Driven By A Vector Field},
  \href{http://prd.aps.org/abstract/PRD/v40/i4/p967_1}{Phys.\ Rev.\ D {\bf 40 } (1989)  967}.

\bibitem{hervik06}
J. D. Barrow and S. Hervik,
  \emph{Anisotropically Inflating Universes},
  Phys. Rev. D \textbf{73} (2006) 023007 [\href{http://arxiv.org/abs/gr-qc/0511127}{arXiv:gr-qc/0511127}].

\bibitem{hervik06nr2}
J. D. Barrow and S. Hervik,
  \emph{On the evolution of universes in quadratic theories of gravity},
  Phys. Rev. D \textbf{74} (2006) 124017 [\href{http://arxiv.org/abs/gr-qc/0610013}{arXiv:gr-qc/0610013}].

\bibitem{carroll07}
L. Ackerman, S. M. Carroll and M. B. Wise, 
  \emph{Imprints of a Primordial Preferred Direction on the Microwave Background},
  Phys.Rev.D \textbf{75} (2007) 083502 [\href{http://arxiv.org/abs/astro-ph/0701357}{arXiv:astro-ph/0701357}].

 \bibitem{golovnev08}
A. Golovnev, V. Mukhanov and V. Vanchurin,
  \emph{Vector Inflation},
 JCAP \textbf{0806} (2008) 009 [\href{http://arxiv.org/abs/0802.2068}{arXiv:astro-ph/0802.2068}].

\bibitem{kanno08}
S. Kanno, M. Kimura, J. Soda and S. Yokoyama,
  \emph{Anisotropic Inflation from Vector Impurity},
JCAP \textbf{0808} (2008) 034 [\href{http://arxiv.org/abs/0806.2422}{arXiv:hep-ph/0806.2422}].  

\bibitem{dav1}
  T.~Koivisto and D.~F.~Mota,
  \emph{Vector Field Models of Inflation and Dark Energy},
  JCAP {\bf 0808} (2008) 021 [\href{http://arxiv.org/abs/0805.4229}{arXiv:astro-ph/0805.4229}].

\bibitem{dav3}
 C.~Germani and A.~Kehagias,
  \emph{P-nflation: generating cosmic Inflation with p-forms},
  JCAP {\bf 0903 } (2009) 028 [\href{http://arxiv.org/abs/0902.3667}{arXiv:astro-ph/0902.3667}].

\bibitem{hervik10}
J. D. Barrow and S. Hervik,
  \emph{Simple Types of Anisotropic Inflation},
  Phys. Rev. D \textbf{81} (2010) 023513 [\href{http://arxiv.org/abs/0911.3805}{arXiv:gr-qc/0911.3805}].

\bibitem{nico2}
  T.~S.~Koivisto and D.~F.~Mota,
  \emph{CMB statistics in noncommutative inflation},
  JHEP {\bf 1102} (2011) 061 [\href{http://arxiv.org/abs/1011.2126}{arXiv:astro-ph/1011.2126}].

\bibitem{dav2}
  B.~Himmetoglu, C.~R.~Contaldi and M.~Peloso,
  \emph{Ghost instabilities of cosmological models with vector fields nonminimally coupled to the curvature},
  Phys.\ Rev.D\  {\bf 80 } (2009) 123530 [\href{http://arxiv.org/abs/0909.3524}{arXiv:astro-ph/0909.3524}].

\bibitem{himmetoglu09}
B. Himmetoglu, C. R. Contaldi and M. Peloso,
  \emph{Instability of anisotropic cosmological solutions supported by vector fields},
Phys.Rev.Lett. \textbf{102} (2009) 111301  [\href{http://arxiv.org/abs/0809.2779}{arXiv:astro-ph/0809.2779}].

\bibitem{himmetoglu08}
B. Himmetoglu, C. R. Contaldi and M. Peloso,
  \emph{Instability of the ACW model, and problems with massive vectors during inflation},
 Phys.\ Rev.D\  {\bf 79 } (2009) 063517 [\href{http://arxiv.org/abs/0812.1231}{arXiv:astro-ph/0812.1231}].

\bibitem{carroll09}
S. M. Carroll, T. R. Dulaney, M. I. Gresham and H. Tam,  
  \emph{Instabilities in the Aether}, Phys.Rev.D \textbf{79} (2009) 065011 [\href{http://arxiv.org/abs/0812.1049}{arXiv:hep-th/0812.1049}] 

\bibitem{soda09}
M. Watanabe, S. Kanno and J. Soda,
  \emph{Inflationary Universe with Anisotropic Hair},
Phys. Rev. Lett. \textbf{102} (2009) 191302 [\href{http://arxiv.org/abs/0902.2833}{arXiv:hep-th/0902.2833}].

\bibitem{soda09mag}
S. Kanno, J. Soda and M. Watanabe,
  \emph{Cosmological Magnetic Fields from Inflation and Backreaction},
  JCAP \textbf{0912} (2009) 009 [\href{http://arxiv.org/abs/0908.3509}{arXiv:astro-ph/0908.3509}].

\bibitem{soda10}
S. Kanno, J. Soda and M. Watanabe,
  \emph{Anisotropic Power-law Inflation},
  JCAP \textbf{1012} (2010) 024 [\href{http://arxiv.org/abs/1010.5307}{arXiv:hep-th/1010.5307}].
  
  \bibitem{soda10pert}
M. Watanabe, S. Kanno and J. Soda,
  \emph{The Nature of Primordial Fluctuations from Anisotropic Inflation},
Prog. Theor. Phys. \textbf{123} (2010) 1041 [\href{http://arxiv.org/abs/1003.0056}{arXiv:astro-ph/1003.0056}].

\bibitem{soda10cmb}
M. Watanabe, S. Kanno and J. Soda,
  \emph{Imprints of anisotropic inflation on the cosmic microwave background},
MNRAS: Letters, \textbf{412} (2011) Issue 1, pp. L83-L87 [\href{http://arxiv.org/abs/1011.3604}{arXiv:astro-ph/1011.3604}].

\bibitem{emami}
 R.~Emami, H.~Firouzjahi, S.~M.~Sadegh Movahed and M.~Zarei,
  \emph{Anisotropic Inflation from Charged Scalar Fields},
 JCAP {\bf 1102}, 005 (2011) [\href{http://arxiv.org/abs/1010.5495}{arXiv:astro-ph/1010.5495}].

\bibitem{emami2}
 R.~Emami and H.~Firouzjahi,
  \emph{Issues on Generating Primordial Anisotropies at the End of Inflation}, \href{http://arxiv.org/abs/1111.1919}{arXiv:astro-ph/1111.1919}. 
 
\bibitem{dimo}
K.~Dimopoulos and J.~M.~Wagstaff,
 \emph{Particle Production of Vector Fields: Scale Invariance is Attractive},
 Phys.\ Rev.\  D {\bf 83} (2011) 023523 [\href{http://arxiv.org/abs/1011.2517}{arXiv:hep-ph/1011.2517}].

\bibitem{dimo2}
K.~Dimopoulos, G.~Lazarides and J.~M.~Wagstaff,
 \emph{Eliminating the $\eta$-problem in SUGRA Hybrid Inflation with Vector Backreaction}, \href{http://arxiv.org/pdf/1111.1929}{arXiv:astro-ph/1111.1929}.

\bibitem{do}
T. Q. Do, W. F. Kao and I. C. Lin,
\emph{Anisotropic power-law inflation for a two scalar fields model},
\href{http://prd.aps.org/abstract/PRD/v83/i12/e123002}{Phys. Rev. D \textbf{83} (2011) 123002}.

\bibitem{wald}
R. M. Wald, 
\emph{Asymptotic behavior of homogeneous cosmological models in the presence of a positive cosmological constant},
\href{http://prd.aps.org/abstract/PRD/v28/i8/p2118_1}{Phys.Rev.D \textbf{28} (1983) 2118}.  

\bibitem{moss}
I. Moss and V. Sahni, 
\emph{Anisotropy In The Chaotic Inflationary Universe},
\href{http://www.sciencedirect.com/science/article/pii/0370269386914887}{Phys. Lett. B \textbf{178} (1986) 159}.

\bibitem{nohairpowerlaw}
Y. Kitada and K. Maeda 
  \emph{Cosmic no-hair theorem in power-law inflation},
  \href{http://prd.aps.org/abstract/PRD/v45/i4/p1416_1}{Phys.Rev.D \textbf{45} (1992) 1416}.  
  
\bibitem{azadeh}
A. Maleknejad, M.M. Sheikh-Jabbari and J. Soda,
\emph{Gauge-flation and Cosmic No-Hair Conjecture}, \href{http://arxiv.org/abs/1109.5573v2}{arXiv: hep-th/1109.5573v2}.

\bibitem{boehmer} C.~G.~Boehmer and D.~F.~Mota,
\emph{CMB Anisotropies and Inflation from Non-Standard Spinors},
  Phys.\ Lett.\ B {\bf 663} (2008) 168 [\href{http://arxiv.org/abs/0710.2003}{arXiv:astro-ph/0710.2003}].

\bibitem{linde88}
A.D. Linde and M.I. Zelnikov,
  \emph{Inflationary universe with fluctuating dimension}, 
 \href{http://adsabs.harvard.edu/abs/1988PhLB..215...59L}{Phys. Lett. B \textbf{215} (1988) 59}.

\bibitem{barrow96}
J. Yearsley and J.D. Barrow,
\emph{Cosmological Models of Dimensional Segregation}, 
\href{http://iopscience.iop.org/0264-9381/13/10/009}{Class. Quantum Grav. \textbf{13} (1996) 2693}.

\bibitem{jacobs68}
K.C. Jacobs,
  \emph{Spatially homogenous and Euclidean cosmological models with shear}, 
  \href{http://adsabs.harvard.edu/abs/1968ApJ...153..661J}{Astrophys. J. \textbf{153} (1968) 661}.

\bibitem{hervik01}
S. Hervik,
\emph{Discrete Symmetries in Translation Invariant Cosmological Models}, 
Gen.Rel.Grav. \textbf{33} (2001) 2027 [\href{http://arxiv.org/abs/gr-qc/0105006}{arXiv:gr-qc/0105006}].

\bibitem{collins77}
  C.B. Collins,
  \emph{Global structure of Kantowski-Sachs cosmological models},
   \href{http://jmp.aip.org/resource/1/jmapaq/v18/i11/p2116_s1}{J.Math.Phys. \textbf{18} (1977) 2116}.  

\bibitem{ellisboka}
  J. Wainwright and G.F.R. Ellis,
  \emph{Dynamical systems in cosmology},
Cambridge University Press, Cambridge - UK (1997).

\bibitem{collins71}
  C.B. Collins and J.M. Stewart,
  \emph{Qualitative cosmology},
   \href{http://adsabs.harvard.edu/abs/1971MNRAS.153..419C}{MNRAS \textbf{153} (1971) 419}.  

\bibitem{tilt1}
S. Hervik, W. C. Lim, P. Sandin and C. Uggla,Ê
\emph{Future asymptotics of tilted Bianchi type II cosmologies},
Class.Quant.Grav. \textbf{27} (2010) 185006 [\href{http://arxiv.org/abs/1004.3661}{arXiv:gr-qc/1004.3661}].

\bibitem{tilt2}
A. Coley and S. Hervik,Ê
\emph{Bianchi models with vorticity: The type III bifurcation},
Class. Quantum Grav. \textbf{25} (2008) 198001 [\href{http://arxiv.org/abs/0802.3629}{arXiv:gr-qc/0802.3629}].

\bibitem{supernova10}
L. Campanelli, P. Cea, G.L. Fogli and A. Marrone,
  \emph{Testing the isotropy of the universe with type Ia supernovae},
  Phys.Rev.D \textbf{83} (2011) 103503 [\href{http://arxiv.org/abs/1012.5596}{arXiv:astro-ph/1012.5596}].

\bibitem{barrow85}
  J.D. Barrow, R. Juszkiewicz and D. H. Sonoda,
  \emph{Universal rotation: how large can it be?},
   \href{http://adsabs.harvard.edu/abs/1985MNRAS.213..917B}{MNRAS \textbf{213} (1985) 917}.  

\bibitem{mota08}
T. Koivisto and D. F. Mota,
  \emph{Accelerating Cosmologies with an Anisotropic Equation of State},
Astrophys.J. \textbf{679} (2008) 1 [\href{http://arxiv.org/abs/0707.0279}{arXiv:astro-ph/0707.0279}].

\bibitem{mota08nr2}
T. Koivisto and D. F. Mota,
  \emph{Anisotropic Dark Energy: Dynamics of Background and Perturbations},
JCAP \textbf{0806} (2008) 018 [\href{http://arxiv.org/abs/0801.3676}{arXiv:astro-ph/0801.3676}].
  
\bibitem{koivi1}
  T.~S.~Koivisto, D.~F.~Mota and C.~Pitrou,
\emph{Inflation from N-Forms and its stability},
 JHEP {\bf 0909} (2009) 092 
  [\href{http://arxiv.org/abs/0903.4158}{arXiv:astro-ph/0903.4158}].
  
\bibitem{boka}
  D. H. Lyth and A. R. Liddle,
  \emph{The Primordial Density Perturbation: Cosmology, Inflation and the Origin of
Structure},
Cambridge University Press, Cambridge - UK (2009). 



\end{thebibliography}
\end{document}